\documentclass[twocolumn,twoside,slac_two]{revtex4}
\usepackage{graphicx}
\usepackage{fancyhdr}
\usepackage{subfigure}
\def\beq{\begin{equation}}
\def\eeq{\end{equation}}
\def\bea{\begin{eqnarray}}
\def\eea{\end{eqnarray}}

\def\DG{\Delta \Gamma_s}

\def\beq{\begin{equation}}
\def\eeq{\end{equation}}
\def\bea{\begin{eqnarray}}
\def\eea{\end{eqnarray}}

\def\dfrac#1#2{{\displaystyle {#1 \over #2}}}
\def\vv{V^*_{cb}V^{\phantom{*}}_{cs}}
\begin{document}

\voffset=-1cm

\title{Lifetime difference in $B_s$ mixing: Standard Model and beyond \footnote{Full consideration of $B_s$ mixing is given in our paper in Phys.Lett.B653:230-240,2007}}
\author{Andriy Badin \vspace{5pt}}
\email{a_badin@wayne.edu}
\affiliation{Department of Physics and Astronomy\\
        Wayne State University, Detroit, MI 48201 \vspace{10pt}}

\author{Fabrizio Gabbiani\vspace{5pt}}
\affiliation{Department of Physics and Astronomy\\
        Wayne State University, Detroit, MI 48201 \vspace{10pt}}

\author{Alexey A.\ Petrov\vspace{5pt}}
\email{apetrov@wayne.edu}
\affiliation{Department of Physics and Astronomy\\
        Wayne State University, Detroit, MI 48201\vspace{5pt}}
\affiliation{Michigan Center for Theoretical Physics\\
University of Michigan, Ann Arbor, MI 48109\vspace{10pt} \\
$\phantom{}$}

\begin{abstract}
\vspace{5pt} We present a calculation of $1/m_b^2$ corrections to
the lifetime differences of $B_s$ mesons in the heavy-quark
expansion. We find that they are small to significantly affect
$\Delta\Gamma$ $(B_s)$ and present the result for lifetime
difference including non-perturbative $1/m_b$ and $1/m_b^2$
corrections. We also analyze the generic $\Delta B$ = 1 New Physics
contributions to the lifetime difference of $B_s$ mesons and provide
several examples

\end{abstract}
\maketitle

Mixing phenomena in heavy bosons system is considered as an
important test of Standard Model(SM) and  a probe for New
Physics(NP) beyond it. Usually it is referred to the fact that such
process occurs only at the one loop level in SM . This makes it
sensitive to the effects of new particles running in the loop. These
interactions induce non-diagonal elements in mass-matrix making
flavor and mass eigenstates to be different. Analysis of mixing in
charm, beauty systems led to positive signals which seem to be very
well explained by Standard Model physics. The lifetime difference
$\Delta\Gamma_s$ is generated by on-mass-shell intermediate states
and seems to be one more test of Standard Model and heavy quark
expansion. Yet some contribution of NP is still possible as an
indirect probe of energy scales beyond currently accessible at
experimental facilities. Further in paper we set up relevant
formalism and discuss the need to compute $1/m_b^2$ corrections.
After these corrections computed we consider impact of New Physics
$\Delta b=1$ interaction on the numerical value of $\Delta\Gamma_s$

\section{Formalism}
The width difference between mass eigenstates is then given
by~\cite{Beneke:1996gn}
\begin{equation}
\label{dgdef} \Delta \Gamma_{B_s}\equiv\Gamma_L-\Gamma_H =
-2\Gamma_{12}=-2\Gamma_{21},
\end{equation}
where $\Gamma_{ij}$ are the elements of the decay-width matrix,
$i,j=1,2$ ($|1\rangle=|B_s\rangle$,
$|2\rangle=|\overline{B_s}\rangle$)

Using optical theorem, off diagonal elements of mixing matrix can be
related to the imaginary part of the forward scattering amplitude:
\begin{eqnarray}
\label{rate} \Gamma_{21}(B_s)&=&\frac{1}{2 M_{B_s}}\langle
\overline{B}_s |{\cal T} | B_s \rangle,\quad\\
 \label{tdef}{\cal T} &=& {\mbox{Im}}~
i \int d^4 x T \left\{ H_{\mbox{\scriptsize eff}}(x)
H_{\mbox{\scriptsize eff}}(0) \right \}.
\end{eqnarray}

where $H_{eff}$ is an effective weak hamiltionian defined as
follows:
\begin{equation}\label{hpeng}
{\cal H}_{eff}=\frac{G_F}{\sqrt{2}}\vv \left(\,\sum^6_{r=1} C_r Q_r
+ C_8 Q_8\right),
\end{equation}

where four-quark operators are defined in the following way:
\begin{eqnarray}\label{q1q2}
Q_1&=& (\bar b_ic_j)_{V-A}(\bar c_js_i)_{V-A},\\
Q_2&=& (\bar b_ic_i)_{V-A}(\bar c_js_j)_{V-A}, \\
\label{q3q4} Q_3&=& (\bar b_is_i)_{V-A}(\bar q_jq_j)_{V-A},\\
Q_4&=& (\bar b_is_j)_{V-A}(\bar q_jq_i)_{V-A},\\
\label{q5q6} Q_5&=& (\bar b_is_i)_{V-A}(\bar q_jq_j)_{V+A},\\
 Q_6&=&(\bar b_is_j)_{V-A}(\bar q_jq_i)_{V+A},\\
\label{q8} Q_8&=& \frac{g}{8\pi^2}m_b\, \bar
b_i\sigma^{\mu\nu}(1-\gamma_5)T^a_{ij} s_j\, G^a_{\mu\nu}.
\end{eqnarray}

In the heavy-quark limit the energy release is large and process is
dominated by short-distance physics. An operator product expansion
can be constructed which results in series of operators suppressed
by powers of $1/m_b$:
\bea
\label{expan}
\Gamma(B_s)_{21}&=&\frac{1}{2 M_{B_s}} \sum_k \langle B_s |{\cal
T}_k | B_s \rangle\nonumber\\
&=&\sum_{k} \frac{C_k(\mu)}{m_b^{k}} \langle B_s |{\cal O}_k^{\Delta
B=2}(\mu) | B_s \rangle.
\eea
 The most recent
calculations of $B_s$ lifetime difference \cite{Beneke:1996gn} and
of QCD corrections to $\Delta\Gamma_s$ \cite{Beneke:1998sy,
Ciuchini:2003ww} do not provide definitive theoretical prediction of
its value. Heavy quark expansion corrections of order of $1/m_b$
appear to be about $25\%$ of leading order and QCD corrections are
as big as $30\%$. \footnote{It was proposed in \cite{Lenz:2006hd}
that four-quark operators governing this interaction can be
redefined in such a way that corrections will be small}. We compute
$1/m_b^2$ corrections in heavy quark expansion to directly check
convergence of this series. In other words we compute matching
coefficients of an effective $\Delta b=2$ lagrangian. Computation of
matrix elements of most of  these operators is rather difficult task
due to lack of results from Lattice QCD and Light cone QCD
calculations. We used a factorization approach to estimate matrix
elements of such operators.

Expanding the operator product (\ref{tdef}) for small $x\sim 1/m_b$,
the transition operator ${\cal T}$ can be written, to leading order
in the $1/m_b$ expansion, as~\cite{Beneke:1996gn,Beneke:1998sy}
\beq\label{tfq} {\cal T}=-\frac{G^2_F m^2_b}{12\pi}(\vv)^2 \, \left[
F(z) Q(\mu_2)+ F_S(z) Q_S(\mu_2) \right], \eeq
which results in~\cite{Ciuchini:2003ww}
\bea\label{tres} &&\Gamma_{21}(B_s)= -\frac{G^2_F m^2_b}{12\pi (2
M_{B_s})}(\vv)^2
\sqrt{1-4z}\times  \nonumber\\
&\times&\left\{\left[(1-z)\,\left(2\, C_1 C_2+N_c
C^2_2\right)+(1-4z) C^2_1/2 \right]\right. \langle Q\rangle\nonumber\\
 &+& \left.(1+2z)\left(2\, C_1
 C_2+N_c C^2_2-C^2_1\right) \langle Q_S \rangle \right\}, \eea
where $z=m_c^2/m_b^2$ and the $\Delta B=2$ operators are as follows:
\bea\label{qqs} Q &=& (\bar b_is_i)_{V-A}(\bar
b_js_j)_{V-A},\nonumber\\ Q_S&=& (\bar b_is_i)_{S-P}(\bar
b_js_j)_{S-P} ~. \eea
Color re-arranged operators $\tilde Q=(\bar b_is_j)_{V-A}(\bar
b_js_i)_{V-A}$ and $\tilde Q_S= (\bar b_is_j)_{S-P}(\bar
b_js_i)_{S-P}$ that appear during calculations were eliminated using
Fiertz identities and equation of motion.

 The Wilson coefficients
$F$ and $F_S$ are obtained by computing the matrix elements of
${\cal T}$ in (\ref{tdef}) 
between
quark states.

The coefficients in the transition operator (\ref{tfq}) at
next-to-leading order, still neglecting the penguin sector, can be
written as~\cite{Beneke:1998sy}:
\bea\label{fz} F(z)&=&F_{11}(z) C^2_2(\mu_1)+ F_{12}(z) C_1(\mu_1)
C_2(\mu_1)+\nonumber\\
&+&F_{22}(z) C^2_1(\mu_1),
\eea
\begin{equation}\label{fij}
F_{ij}(z)=F^{(0)}_{ij}(z)
+\frac{\alpha_s(\mu_1)}{4\pi}F^{(1)}_{ij}(z), \eeq
$F_S(z)$ has similar structure. The leading order functions
$F^{(0)}_{ij}$, $F^{(0)}_{S,ij}$ read explicitly
\bea
\label{f011}
F^{(0)}_{11}(z)&=&3\sqrt{1-4z}(1-z),\nonumber\\
F^{(0)}_{S,11}(z)&=&3\sqrt{1-4z}(1+2z),\\
\label{f012}
F^{(0)}_{12}(z)&=&2\sqrt{1-4z}(1-z)\nonumber\\
F^{(0)}_{S,12}(z)&=&2\sqrt{1-4z}(1+2z),\\
\label{f022}
F^{(0)}_{22}(z)&=&\frac{1}{2}(1-4z)^{3/2}\nonumber\\
F^{(0)}_{S,22}(z)&=&-\sqrt{1-4z}(1+2z).
\eea
The next-to-leading order (NLO) QCD expressions of $F^{(1)}_{ij}$,
$F^{(1)}_{S,ij}$ and corrections to Eq.~(\ref{tfq}) arrising from
penguin diagram are given in Ref.~\cite{Beneke:1998sy}.

\section{$1/m_b^n$ corrections}
The general expression for lifetime difference of $B_s$ mesons can
be presented in the following way: \bea\label{corr}
&&\Gamma_{21}(B_s)
= -\frac{G^2_F m^2_b}{12\pi (2 M_{B_s})}(\vv)^2\times\nonumber\\
&\times&\,\left\{\left[F(z)+P(z)\right]Q+
\left[F_S(z)+P_S(z)\right]Q_S\right.\\
 &+&\left.\delta_{1/m} + \delta_{1/m^2}\right\}\nonumber \eea where $\delta_{1/m}$ and
$\delta_{1/m^2}$ denote contribution from operators suppressed as
$1/m_b$ and $1/m_b^2$ respectively. These terms and their numerical
values are computed further.

The matrix elements for $Q$ and $Q_S$ can be parametrized in the
following way~\cite{Beneke:1996gn,Beneke:1998sy,Ciuchini:2003ww}
\bea
\label{meqs}
\langle \overline B_s\vert Q\vert B_s\rangle &=&
f^2_{B_s}M^2_{B_s}2\left(1+\frac{1}{N_c}\right)B,
\\[0.1cm]
\langle \overline B_s\vert Q_S \vert B_s\rangle &=&
-f^2_{B_s}M^2_{B_s}
\frac{M^2_{B_s}}{(m_b+m_s)^2}\left(2-\frac{1}{N_c}\right)B_S,
\nonumber
\eea
\noindent where $M_{B_s}$ and $f_{B_s}$ are the mass and decay
constant of the $B_s$ meson and $N_c$ is the number of colors. $B$
and $B_S$ are defined such that $B=B_S=1$ corresponds to the
factorization (or `vacuum insertion') approach, which can provide a
first estimate. Their numerical values are known from Lattice QCD
calculations.

The $1/m_b$ corrections are obtained expanding amplitude
Eq.~(\ref{rate}) in terms of light quark momentum and matching it to
four-quark operators that contain derivatives \cite{Beneke:1996gn,
Ciuchini:2003ww},

The $\delta_{1/m}$ term can be written in the following form:
\bea
\label{OurCorrection}
\delta_{1/m} &=& \sqrt{1-4z}\left\{ (1 +
2z)\left[C^2_1\left(R_2 + 2 R_4\right)- \right.\right.\nonumber\\
&-&2\left.(2C_1 C_2+N_c C_2^2)\left(R_1+R_2\right)\right]\frac{z^2}{4}\nonumber \\
&-&\frac{12 z^2}{1 - 4z}\left[(2 C_1 C_2+N_c C^2)\left(R_2+2
R_3\right)\right.\nonumber\\
 &+&\left.\left. 2C^2_1 R_3\right]\right\} \eea
where additional operators that contain derivatives appear
\bea R_1 &=& \dfrac{m_s}{m_b}\bar b_i \gamma^{\mu} (1-\gamma_5) s_i
~\bar b_j \gamma_{\mu} (1+\gamma_5) s_j\nonumber\\
R_2 &=& \dfrac{1}{m^2_b}\bar b_i {\overleftarrow D}_{\!\rho}
\gamma^\mu (1-\gamma_5)\overrightarrow{D}^\rho s_i
~\bar b_j\gamma_\mu(1-\gamma_5)s_j\,,\nonumber \\
R_3 &=& \dfrac{1}{m^2_b}\bar b_i{\overleftarrow D}_{\!\rho}
(1-\gamma_5)\overrightarrow{D}^\rho s_i~\bar b_j(1-\gamma_5)s_j\\
R_4 &=& \dfrac{1}{m_b}\bar b_i(1-\gamma_5)i\overrightarrow{D}_\mu
s_i ~\bar b_j\gamma^\mu(1-\gamma_5)s_j\,. \nonumber\label{rrt1} \eea
Their matrix elements are~\cite{Beneke:1996gn, Ciuchini:2003ww}:
\bea \langle \label{corr1}\overline  B_s | R_1 | B_s \rangle  &=&
\left(2+\frac{1}{N_c}\right)\,\dfrac{m_s}{m_b}\, f_{B_s}^2
M_{B_s}^2\,B^{s}_1\\
 \overline  B_s | R_2 | B_s \rangle
&=& \left(-1+\frac{1}{N_c}\right)\,f_{B_s}^2 M_{B_s}^2
\left( \dfrac{M_{B_s}^2}{m_b^2} - 1 \right)\,B^{s}_2 \nonumber\\
\langle  \overline  B_s | R_3 | B_s \rangle  &=&
\left(1+\frac{1}{2N_c}\right)\,f_{B_s}^2 M_{B_s}^2 \left(
\dfrac{M_{B_s}^2}{m_b^2} - 1 \right)\,B^{s}_3\nonumber\\
 \langle \overline
B_s | R_4 | B_s \rangle &=& -f_{B_s}^2 M_{B_s}^2 \left(
\dfrac{M_{B_s}^2}{m_b^2} - 1 \right)\,B^{s}_4\,.\nonumber \eea
Among these $B$-parameters, $B^{s}_1$ and $B^{s}_2$ are the most
widely studied and well known in lattice  and light cone QCD. In
this paper we use the results of Ref.~\cite{Becirevic:2001xt}. The
rest of "bag" parameters is estimated using a vacuum insertion
approximation. The color-rearranged operators $\widetilde{R}_i$ were
eliminated using Fierz identities and the equations of motion as in
Eq.~(\ref{qqs}).

As it was mentioned earlier $O(1/m_b)$ corrections are quite large
\cite{Beneke:1996gn, Ciuchini:2003ww}. Computing $O(1/m_b^2)$ we
directly control convergence of $1/ m$ expansion in lifetime
difference calculation.  At this order we get more operators that
contribute to the $\Delta\Gamma(B_s)$. There are two different types
of operators. One class of them involves operators computed by
further expansion of Eq.~(\ref{rate}) - they are called kinetic
corrections. Another type arises from interaction of quarks with
background gluon field.

The kinetic corrections can be written as:
\bea\label{OurCorrection1} &&\delta_{1/m^2} = \dfrac{24 z^2}{(1-4
z)^2}(3-10z)\left[ C_1^2 W_3+\right.\nonumber\\
&+&\left.(2\, C_1 C_2 +N_c C_2^2) (W_3+W_2/2)\right]\nonumber \\
&-&\dfrac{12 z^2}{1-4 z}\dfrac{m_s^2}{m_b^2}\left[
C_1^2 Q_S-(2\, C_1 C_2 +N_c C_2^2) (Q_S+Q/2)\right]\nonumber \\
&+&\dfrac{24 z^2}{1-4 z}\left[
2 C_1^2 W_4-2\, (2\, C_1 C_2 +N_c C_2^2) (W_1+W_2/2)\right]\nonumber \\
&-&(1-2 z)\dfrac{m_s^2}{m_b^2}(C_1^2+2\, C_1 C_2 +N_c C_2^2)Q_R.
\eea
 The operators
in Eq.~(\ref{OurCorrection1}) are defined as
\bea
Q_R &=& (\bar b_is_i)_{S+P}(\bar b_js_j)_{S+P}\,,\nonumber \\
W_1 &=& \dfrac{m_s}{m_b}
\bar b_i\overleftarrow{D}^{\alpha}(1-\gamma_5)\overrightarrow{D}_{\alpha}s_i~\bar b_j(1+\gamma_5)s_j\,,\nonumber \\
W_2 &=& \dfrac{1}{m_b^4} \bar
b_i\overleftarrow{D}^{\alpha}\overleftarrow{D}^{\beta}\gamma^{\mu}(1-\gamma_5)\overrightarrow{D}_{\alpha}\overrightarrow{D}_{\beta}s_i
~\bar b_j\gamma_{\mu}(1-\gamma_5)s_j\,,  \nonumber \\
W_3 &=& \dfrac{1}{m_b^4} \bar
b_i\overleftarrow{D}^{\alpha}\overleftarrow{D}^{\beta}(1-\gamma_5)\overrightarrow{D}_{\alpha}\overrightarrow{D}_{\beta}s_i
~\bar b_j (1-\gamma_5) s_j\,, \nonumber \\
W_4 &=& \dfrac{1}{m_b^4} \bar
b_i\overleftarrow{D}^{\alpha}(1-\gamma_5)i
\overrightarrow{D}_{\mu}\overrightarrow{D}_{\alpha}s_i ~\bar
b_j\gamma^{\mu}(1-\gamma_5)s_j\,, \eea
where, as before, we have eliminated the color-rearranged operators
$\widetilde{W}_i$ in favor of the operators $W_i$. Due to absence of
results from lattice and light cone QCD, the parametrization of the
matrix elements of these operators is given. In the  pure
factorization approach all the bag parameters $\alpha_i$  should be
set to 1:
\bea \label{corr2} \langle \overline  B_s |Q_R| B_s \rangle &=&
-f^2_{B_s}M^2_{B_s}
\frac{M^2_{B_s}}{(m_b+m_s)^2}\alpha_1\,,  \\
\langle \overline  B_s |W_1| B_s \rangle &=& \left(1+\frac{1}{2
N_c}\right)\,f_{B_s}^2 M_{B_s}^2
\left( \dfrac{M_{B_s}^2}{m_b^2} - 1 \right)\alpha_2\,, \nonumber \\
\langle \overline  B_s |W_2| B_s \rangle &=&
\frac{1}{2}\left(-1+\frac{1}{N_c}\right)\,f_{B_s}^2 M_{B_s}^2
\left( \dfrac{M_{B_s}^2}{m_b^2} - 1 \right)^2\alpha_3\,, \nonumber \\
\langle \overline  B_s |W_3| B_s \rangle &=& \frac{1}{2}
\left(1+\frac{1}{2 N_c}\right)\,f_{B_s}^2 M_{B_s}^2
\left( \dfrac{M_{B_s}^2}{m_b^2} - 1 \right)^2\alpha_4\,, \nonumber \\
\langle \overline  B_s |W_4| B_s \rangle &=& -\frac{1}{2}\,f_{B_s}^2
M_{B_s}^2 \left( \dfrac{M_{B_s}^2}{m_b^2} - 1 \right)^2\alpha_5\,.
\nonumber \eea
\begin{figure}[h]
\centering
\includegraphics[width=80mm]{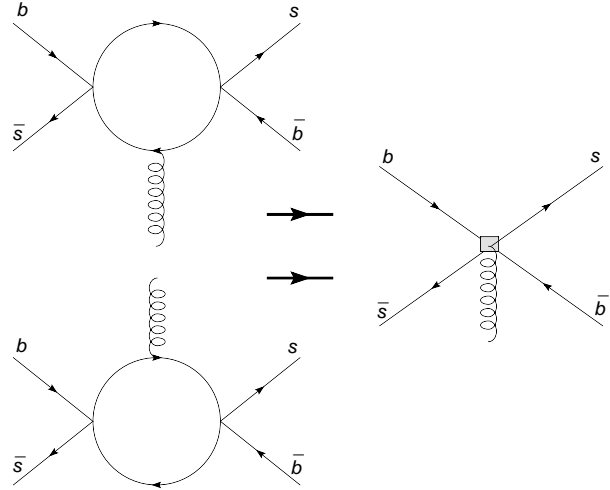}
\caption{Diagrams contributing to corrections due to interaction
with background gluon field.} \label{gluon}
\end{figure}
In addition to the set of kinetic corrections considered above, the
effects of the interactions of the intermediate quarks with
background gluon fields should also be included at this order. The
contribution of those operators can be computed from the diagram on
Fig.\ref{gluon}, resulting in
\bea &&{\cal T}_{{\rm spec}, G} = -
\dfrac{G_F^2(\vv)^2}{4\pi\sqrt{1-4z}} \left\{C_1^2\left[(1-4z)
P_1 P_2+\right.\right.\nonumber\\
&-&(1-4z)+\left.4z P_3-4z P_4 \right]\nonumber\\
&+& \left. 4\; C_1 C_2 z\left[P_5+P_6-P_7-P_{8}\right]\right\}.
\nonumber \eea
The local four-quark operators contributing to this correction are
given in  Eq.~(\ref{poper}).
\bea \label{poper} P_1 &=& \bar
b_i\gamma^{\mu}(1-\gamma_5)s^{\phantom{l}}_i~ \bar
b^{\phantom{l}}_k\gamma^{\nu}(1-\gamma_5)t_{kl}^a\widetilde{G}_{\mu\nu}^a
s^{\phantom{l}}_l\\
P_2 &=& \bar
b_k\gamma^{\mu}(1-\gamma_5)t_{kl}^a\widetilde{G}_{\mu\nu}^a s_l~
\bar b^{\phantom{l}}_i\gamma^{\nu}(1-\gamma_5)s^{\phantom{l}}_i\,,  \nonumber\\
P_3 &=& \dfrac{1}{m_b^2}\bar
b_i\overleftarrow{D}^{\mu}\overleftarrow{D}^{\alpha}\gamma^{\alpha}(1-\gamma_5)
s^{\phantom{l}}_i~ \bar
b^{\phantom{l}}_k\gamma_{\nu}(1-\gamma_5)t_{kl}^a\widetilde{G}_{\mu\nu}^a
s^{\phantom{l}}_l\,,
\nonumber\\
P_4 &=& \dfrac{1}{m_b^2}\bar
b_k\overleftarrow{D}^{\nu}\overleftarrow{D}^{\alpha}\gamma^{\mu}(1-\gamma_5)
t_{kl}^a\widetilde{G}_{\mu\nu}^a s^{\phantom{l}}_l
~\bar b^{\phantom{l}}_i\gamma_{\alpha}(1-\gamma_5)s^{\phantom{l}}_i\,, \nonumber \\
P_5 &=& \dfrac{1}{m_b^2}\bar
b_k\overleftarrow{D}^{\nu}\overleftarrow{D}^{\alpha}\gamma^{\mu}(1-\gamma_5)
s^{\phantom{l}}_i~t_{kl}^a\widetilde{G}_{\mu\nu}^a
~\bar b^{\phantom{l}}_i\gamma_{\alpha}(1-\gamma_5)s^{\phantom{l}}_l\,, \nonumber \\
P_6 &=& \dfrac{1}{m_b^2}\bar
b_i\overleftarrow{D}^{\nu}\overleftarrow{D}^{\alpha}\gamma^{\mu}(1-\gamma_5)
s^{\phantom{l}}_k~t_{kl}^a\widetilde{G}_{\mu\nu}^a ~\bar
b^{\phantom{l}}_l\gamma_{\alpha}(1-\gamma_5) s^{\phantom{l}}_i\,,
\nonumber \\
P_7 &=& \dfrac{1}{m_b^2}\bar
b_k\overleftarrow{D}^{\mu}\overleftarrow{D}^{\alpha}\gamma^{\alpha}(1-\gamma_5)
s^{\phantom{l}}_i~t_{kl}^a\widetilde{G}_{\mu\nu}^a ~\bar
b^{\phantom{l}}_i\gamma_{\nu}(1-\gamma_5)s^{\phantom{l}}_l\,,
\nonumber \\
P_8 &=& \dfrac{1}{m_b^2}\bar
b_i\overleftarrow{D}^{\mu}\overleftarrow{D}^{\alpha}\gamma^{\alpha}(1-\gamma_5)
s^{\phantom{l}}_k~t_{kl}^a\widetilde{G}_{\mu\nu}^a ~\bar
b^{\phantom{l}}_l\gamma_{\nu}(1-\gamma_5) s^{\phantom{l}}_i.
\nonumber
\eea
Following \cite{Gabbiani:2004tp} these operators are parametrized
the following way:
\beq \label{glucorr} \langle B_s | P_i | B_s \rangle = \dfrac{1}{4}
f^2_{B_{s}} M^2_{B_{s}}
\left(\dfrac{M_{B_s}^2}{m^2_b}-1\right)^2\beta_i. \eeq
It is hard to obtain precise prediction for lifetime difference with
so many operators contributing. Nevertheless contribution from
$\delta_{1/m}$ and $\delta_{1/m^2}$ can be evaluated.  In our
numerical calculations we assume the pole mass of b-quark to be
$m_b=4.8 \pm 0.2 GeV$ and $f_B =  230 \pm 25 MeV$. In order to see
the effect of $O(1/m_b^2)$ corrections we fix all perturbative
parameters in the middle of their allowed ranges to show dependence
of $\Delta\Gamma_{B_s}$ on non perturbative parameters $B_i,\
\alpha_i,\ \beta_i$ defined in Eq.\ref{meqs}, \ref{corr1},
\ref{corr2}, \ref{glucorr}
\begin{eqnarray}
&&\Delta\Gamma_{B_s}=\left[ 0.0005 B + 0.1732 B_s + 0.0024 B_1\right.\nonumber\\
&&- 0.0237 B_2-0.0024B_3-0.0436B_4\nonumber\\
&&+2\times10^{-5} \alpha_1+4\times10^{-5} \alpha_2 + 4\times10^{-5} \alpha_3 \nonumber\\
&&+0.0009\alpha_4  -0.0007\alpha5\nonumber\\
&&+0.0002\beta_1- 0.0002\beta_2+6\times10^{-5}\beta_3\\
&&-6\times10^{-5}\beta_4-1\times10^{-5}\beta_5\nonumber\\
&&\left.-1\times10^{-5}\beta_6+1\times10^{-5}\beta_7 +
1\times10^{-5}\beta_8\right] (ps^{-1})\nonumber
\end{eqnarray}
It is obvious that $O(1/m_b^2)$ corrections provide minor effect on
calculation of $B_s - \overline{B}_s$ lifetime difference.
Contribution from interaction with background gluon field is
essentially negligible.

To obtain full SM prediction of $\Delta\Gamma_{B_s}$ we vary values
of parameters of matrix elements. We generate 100000-point
probability distribution of the lifetime difference obtained by
randomly varying our parameters within $\pm 30\%$ range around their
factorization value  or within $\pm 1 \sigma$ for parameters known
from experimental data or lattice QCD calculations. The resulting
distribution is presented on the Fig.\ref{fig:result}
\begin{figure}[h]
\centering
\includegraphics[width=80mm]{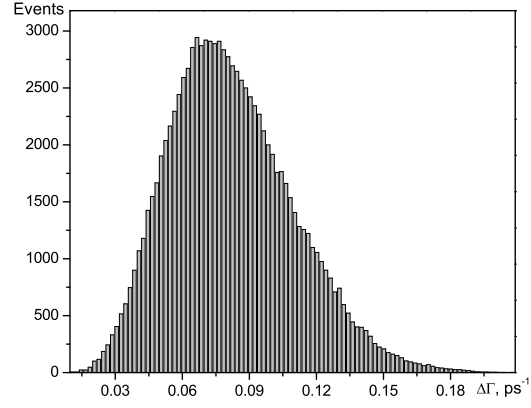}
\caption{Histogram showing the random distribution around the
central values of various parameters contributing to $B_s$-lifetime
difference $\Delta\Gamma_{B_s}$.} \label{fig:result}
\end{figure}
There is no theoretically consistent way to treat this diagram since
it is not expected for theoretical predictions to have Gaussian
distribution. Nevertheless we can give a numerical prediction
estimating position of peak as the most probable value and the peak
width at half of height as theoretical uncertainty.
\bea \label{main_result} \Delta\Gamma_{B_s}&=&0.072\pm{\displaystyle
{0.034 \over 0.030}}
ps^{-1}\nonumber\\
\frac{\Delta\Gamma_{B_s}}{\Gamma_{B_s}}&=&0.104\pm0.049 \eea
where in latter result we added theoretical error for our
calculation of $\Delta\Gamma_{B_s}$ and experimental error from
determination of $\Gamma_{B_s}$ in quadrature. Additional
improvement in lattice or QCD sum rules determination of "bag"
parameters would make this prediction even more solid.

\section{New Physics contributions to lifetime difference}

In the previous section it was shown that $O(1/m_b^2)$corrections to
the lifetime difference of $B_s$ and $\overline{B}_s$ mesons are
small. Since we have reliable prediction of $\Delta\Gamma_{B_s}$ it
might be interesting to consider possible effects of physics beyond
the Standard Model on the lifetime difference in $B_s$ system.

As was pointed out long time
ago~\cite{Grossman:1996er,Dunietz:2000cr}, CP-violating
contributions to $M_{12}$ must {\it reduce} the lifetime diffence in
$B_s$-system, as
\beq \DG = \DG^{SM} \cos^2 2 \theta_s, \eeq
where $\theta_s$ is a CP-violating phase of $M_{12}$, which is
thought to be dominated by some $\Delta B =2$ New Physics. On other
hand, CP conserving  $\Delta B=1$ NP amplitude can interfere with SM
contribution constructively or destructively, depending on the NP
model.

There was no spectacular NP phases observed in $B_s$ mixing, thus it
is important to estimate the CP-conserving contribution to $\DG$. We
shall consider it using the generic set of effective operators, and
then apply our results to popular extensions of the SM.

Using the completeness relation the NP contribution to the
$B^0_s$-${\overline B}^0_s$ lifetime difference becomes
\begin{eqnarray}\label{gammaope}
y \ &=& \  \frac{2}{M_{\rm B_s} \Gamma_{\rm B_s}}\, \langle
\overline{B}_s |
    {\rm Im}\, {\cal T} | B_s \rangle \ \ ,
\\
{\cal T} \ &=& \
    \,i\! \int\! {\rm d}^4 x\, T \left(
    {\cal H}^{\Delta b=-1}_{SM} (x)\, {\cal H}^{\Delta b=-1}_{NP}(0)
\right) \ \ .\nonumber
\end{eqnarray}
We represent the generic NP $\Delta b=1$ hamiltonian as
\bea\label{HamNP} && {\cal H}^{\Delta B=-1}_{NP} = \sum_{q,q'} \
D_{qq'} \left[\overline {\cal C}_1(\mu) Q_1 + \overline {\cal C}_2
(\mu) Q_2 \right]\ ,
\\
&& Q_1 = \overline{b}_i \overline\Gamma_1 q_j' ~ \overline{q}_j
\overline\Gamma_2 s_i \ , \ \ Q_2 = \overline{b}_i \overline\Gamma_1
q_i' ~ \overline{q}_j \overline\Gamma_2 s_j\ , \nonumber \eea
where $\overline\Gamma_{1,2}$ are arbitrary combinations of Dirac
matrices and $\overline {\cal C}_{1,2}(\mu)$ are Wilson coefficients
evaluated at energy scale $\mu$. This gives us the following
contribution to lifetime difference:
\begin{eqnarray}
\label{New Physics result} &&\Delta\Gamma_{NP}= \frac{4G_F
\sqrt{2}}{M_{B_s}}\sum_{qq'}D_{qq'}V^{\ast}_{qb}V_{q's}\\
&\times&\left(K_1\delta_{i\beta}\delta_{k\gamma}+
K_2\delta_{k\beta}\delta_{i\gamma}\right)\sum_{j=1}^{5}I_j(x,x')
\langle\overline{B_s}|O_j^{i\beta k\gamma}|B_s\rangle\nonumber
\end{eqnarray}
where $i, \beta, g, \gamma$ stand for color indices, operators
$O_j^{i\beta k\gamma}$ are the following:
\begin{eqnarray}
O_1^{i\beta k\gamma}&=&\left(\bar{b}_i\Gamma^{\nu}\gamma^{\rho}
\Gamma_2s_{\gamma}\right)\left(\bar{b}_k\Gamma_{1}
\gamma_{\rho}\Gamma_{\nu}s_{\beta}\right)\nonumber \\
O_2^{i\beta
k\gamma}&=&\left(\bar{b}_i\Gamma^{\nu}\hat{p}\Gamma_2s_{\gamma}\right)
\left(\bar{b}_k\Gamma_{1}\hat{p}\Gamma_{\nu}s_{\beta}\right)\nonumber\\
\label{New Physics operators} O_3^{i\beta
k\gamma}&=&\left(\bar{b}_i\Gamma^{\nu}\Gamma_2s_{\gamma}\right)
\left(\bar{b}_k\Gamma_{1}\hat{p}\Gamma_{\nu}s_{\beta}\right)\\
O_4^{i\beta
k\gamma}&=&\left(\bar{b}_i\Gamma^{\nu}\hat{p}\Gamma_2s_{\gamma}
\right)\left(\bar{b}_k\Gamma_{1}\Gamma_{\nu}s_{\beta}\right)\nonumber\\
O_5^{i\beta
k\gamma}&=&\left(\bar{b}_i\Gamma^{\nu}\Gamma_2s_{\gamma}\right)
\left(\bar{b}_k\Gamma_{1}\Gamma_{\nu}s_{\beta}\right)\nonumber,
\end{eqnarray}
with $p$ being a $b$-quark momentum, and $K_i$ are the following
combinations of Wilson coefficients
\begin{eqnarray}
K_1&=&(C_2\overline{C}_2 N_c + (C_2\overline{C}_1 +
C_1\overline{C}_2))\nonumber\\
K_2&=&C_1\overline{C}_1
\end{eqnarray}
with the number of colors $N_c$=3.

 Defining $z\equiv m_q^2/m_b^2$
and $z'\equiv m_q'^2/m_b^2$ coefficients $I_j(z,z')$ can be written
as follows:
\begin{eqnarray}
I_1(z,z')&=&-\frac{\Phi
m_c}{48\pi}\left[1-2(z+z')+(z-z')^2\right]\nonumber\\
I_2(z,z')&=&-\frac{\Phi}{24m_c\pi}\left[1+(z+z')-2(z-z')^2\right]\nonumber\\
I_3(z,z')&=&\frac{\Phi }{8\pi}\sqrt{z}\left[1+z'-z\right]\\
I_4(z,z')&=&-\frac{\Phi }{8\pi}\sqrt{z'}\left[1-z'+z\right]\nonumber\\
I_5(z,z')&=&\frac{\Phi m_c}{4\pi}\sqrt{zz'}\nonumber,
\end{eqnarray}
where $\Phi$ is available phase space of process
$\Phi=m_c/2\left(1-2(z+z')+(z-z')^2\right)^{1/2}$

\subsection{Multi-Higgs model}\label{Multi-Higgs}
One of possible realizations of New Physics is a charged Higgs
doublet model proposed in \cite{Golowich:1979hd}. This model
provides new flavor changing interaction mediated by charged Higgs
bosons. It leads to the following four-fermion interaction:
\begin{equation}
\label{Higgs doublet hamiltonian} {\cal{H}}^{\Delta
B=1}_{ChH}=-\frac{\sqrt{2}G_F}{M_H^2}\ \overline{b}_i
\overline\Gamma_1 q_i' ~ \overline{q}_j \overline\Gamma_2 s_j\,
\end{equation}
where $\overline{\Gamma}_i, \ i=1,2$ are
\begin{eqnarray}
\overline\Gamma_1 &=& m_bV_{cb}^{\ast}\cot\beta P_L - m_cV_{cb}^{\ast}\tan\beta P_R\nonumber\\
\overline\Gamma_2 &=& m_sV_{cs}\cot\beta P_R - m_cV_{cs}\tan\beta
P_L
\end{eqnarray}
This leads to three operators with various coefficients, matrix
elements of which contribute to the $y_{ChH}$:
\begin{eqnarray}
&y_{ChH}=\frac{8G_F^2m_b^2}{M_B\Gamma_B}\frac{(V_{cb}^{\ast}V_{cs})^2}{M_H^2}\times\nonumber\\
&\left[\langle Q_1\rangle\left(4K_2x_sI_1\cot^2\beta+2(\cot^2\beta m_b^2x_sI_2-m_bxI_4)(K_2-K_1)\right)\right.+\nonumber\\
&\left.+\langle Q_2\rangle\left(-2K_1x_sI_1\cot^2\beta+(\cot^2\beta
m_b^2x_sI_2-m_bxI_4)(K_2-K_1)\right)+\right.\nonumber\\
&\left.+\langle Q_3\rangle(K_1+K_2)\left(x^2\tan^2\beta
I_5-m_bxI_3\right)\right]
\end{eqnarray}
$I_i$ and $K_i$ were defined above. $x=m_c/m_b$ and $x_s=m_s/m_b$.
$\langle Q_i\rangle$ are as follows:
\begin{eqnarray}
Q_1 &=& (\overline b_i)_L (s_i)_R (\overline b_k)_R ( s_k)_L \\
\langle Q_1\rangle
&=&-\frac{1}{4}f_B^2M_B^2\frac{M_B^2}{(m_b+m_s)^2}
\left(2+\frac{1}{N_c}\right)
\nonumber\\
Q_2 &=& (\overline b_i)_R\gamma^{\nu} (s_i)_R (\overline
b_k)_L\gamma_{\nu} (s_k)_L \\
\langle Q_2\rangle &=&
-\frac{1}{2}f_B^2M_B^2\left(1+\frac{2}{N_c}\right)\nonumber\\
Q_3 &=& (\overline b_i)_L\gamma^{\nu} (s_i)_L (\overline
b_k)_L\gamma_{\nu} (s_k)_L \\
\langle Q_3\rangle
&=&\frac{1}{2}f_B^2M_B^2\left(1+\frac{1}{N_c}\right) \nonumber
\end{eqnarray}
For values of $M_H = 85GeV$ and $\cot\beta=0.05$ \cite{PDG} it gives
$y_{ChH}\approx 0.0034$. This is about 10\% of Standard Model value.
Dependence of $y_{ChH}$ on mass of Higgs boson is given on
Fig.\ref{fig:Higgs}
\begin{figure}[h]
\centering
\includegraphics[width=80mm]{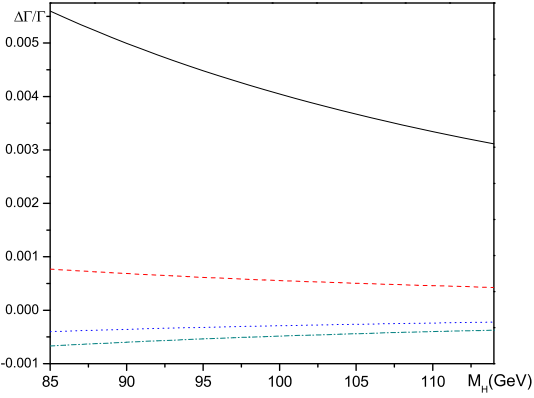}
\caption{Dependence of $y_{ChH}$  on mass of Higgs boson: solid line
- $\tan\beta=20$,dashed line - $\tan\beta=10$, dotted line -
$\tan\beta=5$, dash-dotted line - $\tan\beta=3$} \label{fig:Higgs}
\end{figure}

\subsection{Left-Right Models}
One of the possible extensions of the SM is a Left-Right Symmetric
Model (LRSM) which assumes the extended $SU(2)_L \times SU(2)_R$
symmetry of the theory. In this model additional flavor changing
interaction is provided by mediating right-handed $W^{(R)}$-bosons.
In this case flavor mixing is described by right-handed CKM matrix
$V_{ik}^{(R)}$ and
\begin{eqnarray}
\overline \Gamma_{1,2}&=&\gamma^{\mu}P_R\\
D_{qq'}&=&V_{cb}^{\ast(R)}V_{cs}^{(R)}\frac{G_F^{(R)}}{\sqrt 2}
\end{eqnarray}
here $\frac{G_F^{(R)}}{\sqrt 2}=g_R^2/8M_{W^(R)}^2$ and for future
calculations we take $g_L=k g_R$

Such model gives us the following prediction for value of $y$:
\begin{eqnarray}
y_{LR}&=&-V_{cb}^{\ast}V_{cs}V_{cb}^{\ast(R)}V_{cs}^{(R)}\
\frac{G_F^2m_b^2 x}{\pi
M_B\Gamma_B}\left(\frac{M_W}{M_W^{(R)}}\right)^2\nonumber\\
&\times&\left[C_1 \langle Q_2\rangle + C_2 \langle
\tilde{Q_2}\rangle\right]
\end{eqnarray}
One of possible realizations of such scenario which gives the
biggest numerical value of $y_{LR}$ is a ''Non-manifest LR``
($V_{ij}^{(R)}\approx 1$) with $M_{W^{(R)}}=1\ TeV$ value of
$y_{LR}\approx -0.015$ was obtained. In case of ''manifest LR``
(($V_{ij}^{(R)}= V_{ij}$)) contribution from this model is less.
Dependence of $y_{LR}$ on mass of $W^{(R)}$ boson is given on Fig.
\ref{fig:LRSM}

\begin{figure}
\centering
\subfigure[Solid line:non manifest LRSM ($k=1$), dashed line: manifest LRSM] 
{
    \label{fig:sub:a}
    \includegraphics[width=8cm]{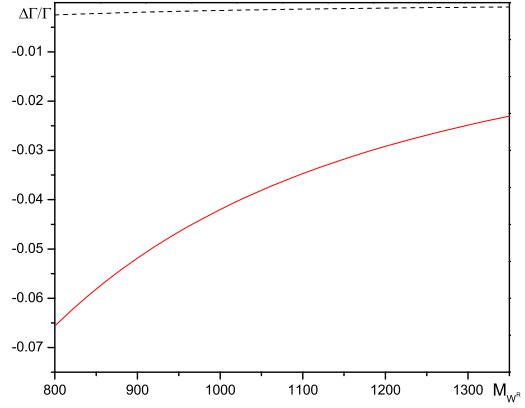}
} \hspace{1cm}
\subfigure[Solid line: $k=1$, dashed line: $k=1.5$, dotted line: $k=2$ ] 
{
    \label{fig:sub:b}
    \includegraphics[width=8cm]{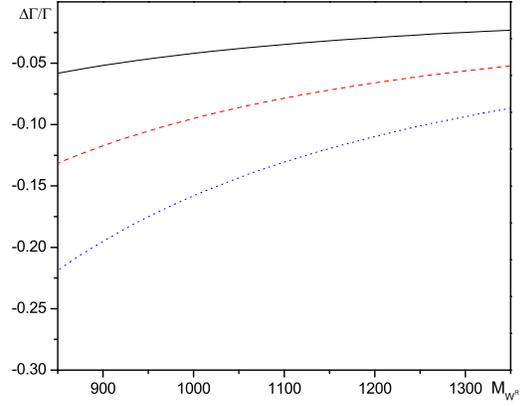}
} \caption{Contribution to $\Delta\Gamma_{B_s}/\Gamma_{B_s}$ in the
left-right symmetric models. (a) - Dependence on $M_W^{(R)}$. (b) -
Dependence on $M_W^{(R)}$ in non-manifest LRSM.}
\label{fig:LRSM} 
\end{figure}

\section{Conclusions}
We computed the subleading $O(1/m_b^2)$ corrections to the lifetime
difference of $B_s$ mesons. The corrections depend on 13
non-perturbative parameters $\alpha_i$ and $\beta_i$. We generated
probability distribution of lifetime difference by varying
parameters $\pm30\%$ around their "factorization" values or within
$1\sigma$ for parameters know from Lattice QCD. The results are
presented on Fig.\ref{fig:result}. Translating this diagram into
numerical prediction  for $\Delta\Gamma_{B_s}/\Gamma_{B_s}$ we
obtained the most precise available today theoretical prediction for
lifetime difference:

\bea \Delta\Gamma_{B_s}&=&0.072\pm{\displaystyle {0.034 \over
0.030}}
ps^{-1}\nonumber\\
\frac{\Delta\Gamma_{B_s}}{\Gamma_{B_s}}&=&0.104\pm0.049
\eea

The effect of $1/m_b^2$ corrections to the lifetime difference is
small.

The generic $\Delta B=1$ New Physics contribution to the lifetime
differnec in $B_s$ system is considered. We considered four-fermion
effective Hamiltionan of the generic Standard Model extension and
computed its contribution to the $\Delta \Gamma_{B_s}$. It can
reduce or increase the SM contribution depending or particular
choice of the model. Two models of physics beyond the Standard Model
considered. Contribution of charged Higgses to the lifetime
difference is negligible. LRSM contribution is significant and
parameters of this model can be constrained based on $\Delta
\Gamma_{B_s}$ measurements.

\end{document}